**Interferometry and Synthesis in Radio Astronomy**
by A. Richard Thompson, James M. Moran and George W. Swenson Jr.

Third Edition
Astronomy and Astrophysics Library
Springer International Publishing
Open Access – 2017, 872 pages
ISBN 978-3-319-44429-1 (hardcover) € 55. -
ISBN 978-3-319-44431-4 (eBook) free

When we look at the sky with the unaided eye its angular resolving power of about one arcminute allows us to see many hundreds of separate stars. We can also see structure in the brightness distribution of the full Moon. A telescope of only 10 cm diameter already provides us with a resolution of about one arcsecond at optical wavelengths. In the early 1940s Grote Reber made the first systematic radio observations of the sky with a parabolic dish of 10 m diameter at a wavelength of about 1.9 m. It resulted in an intensity map of the "radio sky" with an angular resolution of about 12 degrees! Although some regions of higher intensity protruded from the smooth radiation from the Galactic Plane, it was out of the question to learn about their character such as determining the celestial position and angular size. Any comparison with an optical companion was impossible.

In this frustrating situation the early radio astronomers, many of whom came from wartime radar laboratories, notably in Australia and England, quickly turned to the use of interferometers in order to improve the angular resolution that would be proportional to the distance between the two elements of the interferometer rather than their individual size. In early experiments in Australia a "sea-cliff" interferometer was realised with a single antenna located about 100 m above sea level using the direct and reflected path from the ocean surface to create interference fringes. The improved resolution led to the optical identification of the powerful radio sources Cygnus A and Taurus A with a distant galaxy and a supernova in our Galaxy, respectively. In the early 1950s several large interferometric radio telescopes were built that operated at relatively long wavelengths of the order of one meter. Both in Australia and in England the method of aperture synthesis was developed in which data from a number of interferometers with different spacing were combined to *synthesise* a two-dimensional telescope.
With the advent of fully steerable parabolic reflectors operating at shorter cm-wavelengths the method of *earth rotation synthesis* was introduced. The object under observation was now tracked for up to 12 hours along its daily path. This causes the projection of the EW-interferometer baseline to rotate with respect to the source, thus providing baselines in all orientations. From these observations a two-dimensional picture of the brightness distribution of the source can be derived. Major early instruments of this category were the "one mile" and "5 km" arrays in Cambridge, UK (1962, 1968), the Westerbork Synthesis Radio Telescopes (1970) and the Very Large Array in the USA (1980). Parallel to these the technique of VLBI (very long baseline interferometry) was developed in the late 1960s.

This all provided sufficient impetus to the authors of the book under review here to present in 1986 a comprehensive discussion of the principles and applications of "Interferometry and Synthesis in Radio Astronomy". It has been the standard text ever since not the least due to a greatly updated second edition in 2001. Now, again 15 years later, the authors have added another 200 pages for a total of almost 900 to cover the developments in the field to the present. I would classify the new edition as the Radio Astronomy equivalent to "Born & Wolf" for Optics. It is complete, almost encyclopaedic. But it is not an encyclopaedia! It is a successful combination of a graduate course textbook and a compendium of answers to any question one might have on the subject. The latter is helped by the systematic reference to original publications and the extensive subject index.
All aspects of the subject are covered in 17 well-structured chapters. The basic concepts are occasionally introduced in a heuristic approach with the rigorous treatment shifted to a later chapter. There is a parallel here with the structure of Born and Wolf's classic book on Optics. While the pointing to later details may occasionally be somewhat frustrating to the expert, it keeps the general reader on track in gaining understanding without getting immersed in fine details. The first chapter provides a very nice historical introduction with abundant use of illustrations from original publications. Contemporary students may be surprised to see how much information could be gleaned from a tracing on an analogue chart recorder. The theory of interferometry and synthesis imaging is presented in chapter 2 and further developed in chapter 3. The essential role of the Fourier Transform and the concept of convolution are discussed in detail. Chapter 4 covers the aspects of the geometry of the interferometer. It describes the sampling of the spatial frequency plane that delivers the visibility function. After a discussion of polarisation the chapter concludes with the introduction of the *measurement equation*. This equation includes all instrumental and outside (e.g. atmosphere) parameters that influence the measured visibility and hence need to be calibrated to obtain intrinsic visibility data on the observed object.

Multi element interferometer systems are usually called (*aperture*) *synthesis arrays* and can be configured geometrically in several ways. These are discussed in chapter 5 with emphasis on practical matters such as selecting the best configuration for a particular observational purpose. The vehicle for this is the *spatial transfer function* that connects the measured visibility function with the object's brightness distribution via the Fourier Transform. Several implementations of actual arrays are described.

Chapter 6 deals with the response of the electronic receiver system. Frequency conversion (mixing) is treated for single- and double-sideband operation along with the methods of delay correction and fringe rotation. The response to noise and the effects of bandwidth are analysed in detail. A chapter on system design follows in which a detailed exposé is presented of receiver and local oscillator systems, including aspects of noise and delay and gain errors on the sensitivity of the system.

We have now reached the point where normally the signals are digitised and chapter 8 on digital signal processing has grown from 50 pages in the second edition to 80 pages in the new edition. It reflects the growing application of digital circuits in the steps from signal reception by the receiver frontend to the digital data that represent the image of the observed object. This chapter treats in detail the sampling and quantisation issues in digital processing. It is followed by a description of digital delay lines and digital correlators that are now standard in synthesis telescopes. The different types of correlators (XF, FX, hybrid and digital) are compared quantitatively.

In 1967 a special form of interferometer was introduced, whereby the physical connection of the elements by cable or radio link was replaced by adding very accurate time stamps to the received signal that were "lined up" in the off-line correlator to create the interference function.
This enabled baselines of hundreds to thousands of kilometers with commensurate extremely high angular resolution. The method is known as Very Long Baseline Interferometry (VLBI). The special features of this technique are covered in chapter 9. It covers the transition from magnetic tape to hard disks for the recording medium that occurred in 2001 at about the time that the second edition was published. Apart from an indication in the Preface, this edition does not mention the further step of using the Internet for data and timing transport that renders the VLBI essentially into a real-time connected interferometer. First experiments along this line have recently been performed with success. Geodesists and astrometrists adopted the VLBI technique with its exquisitely high angular resolution for the precise measurement of continental drift and the position of celestial objects. The great advances obtained in this field are presented in chapter 12.

Chapters 10 on Calibration and Imaging and 11 on Further Imaging Techniques are more than double the size from the second edition, reflecting the enormous development of these techniques over the last 15 years. In addition to the full Fourier Transformation of the visibility data to obtain the source brightness image, the use of model fitting and closure phase is widely used. These are given extensive attention in this edition. Image processing algorithms, such as CLEAN and self-calibration schemes, are discussed and also here the advances in the field are striking.

A serious impediment to interferometry with spatially separated elements is the influence of the fluctuating index of refraction of the propagation medium, such as the earth's ionosphere and troposphere. The resulting path length variations cause differential phase errors in the measured complex visibilities that are hard, if at all, to separate from the true celestial signal. Also in this area great progress in understanding and new methods of measurement has been achieved since the publication of the second edition. In the current version separate chapters have been assigned to the treatment of the neutral atmosphere (troposphere) in chapter 13 and an ionised medium (ionosphere, interplanetary and interstellar) in chapter 14. The emergence of large arrays with baselines up to 10 km or more (for instance ALMA in the Atacama Desert of Chile) has inspired strong efforts toward measuring the tropospheric pathlength variations (mainly caused by water vapour) in real time to correct the observed visibility phase. Thus separate small antennas measure the water vapour content along the line of sight from which the path length variations can be derived.

Such measurements are routinely made in the evaluation of the suitability of a site for a millimeter or submillimeter wavelength telescope. Where the troposphere seriously affects observations at short cm and mm wavelengths, long wavelength observations suffer under the influence of variations in the ionosphere. The emergence of large arrays at meter wavelengths (for instance LOFAR) puts the need for corrections on the front burner. These telescopes also have to cope with the generally unavoidable interference from man-made radiation, be it TV transmitters, communication channels or industrial "noise". Chapter 16 is devoted to this mundane but important aspect of observational radio astronomy.

In chapter 15 the authors return to the theoretical basis of synthesis in a full discussion of the van Cittert-Zernike theorem and aspects of coherence of the radiation field. This chapter can be considered the rigorous treatment of the material presented in chapters 3 and 4. The book closes with a chapter on "Related Techniques" by shortly discussing the intensity interferometer, lunar occultation methods, tracking of space debris and optical interferometry.

Most chapters have appendices with full mathematical derivations and additional material and an extensive index of names and subjects is added. Throughout the book references to original papers are given and the resulting reference lists should make the use of Google Scholar superfluous.

Overall the book delivers a comprehensive treatment of all aspects of interferometry and synthesis in a lucid and flowing tale that is a pleasure to read. The historical notes that are scattered throughout the text add flavour to the reading. My favourite is the discovery at the Parkes telescope that some weird intermittent interference was caused by a microwave oven in one of the observatory buildings. For all working in this field the book is invaluable. So, what does it cost? Here is the best of all: it is published as Open Access. It can be downloaded for free by anyone who is interested in an 880-page masterpiece. For those, as your reviewer, who prefer to hold this "brick" in their hands, Springer will sell it to you for about 55 Euros, a remarkable small amount, certainly for Springer.

Jacob W. M. Baars
Max-Planck-Institut für Radioastronomie
Bonn, Germany